\documentclass[slac_one]{revtex4}
\usepackage{graphicx}
\usepackage{fancyhdr}
\usepackage{bm}

\begin{document}

\title{Baryon Spectroscopy from Lattice QCD} 

%

\author{Colin Morningstar (for the Hadron Spectrum Collaboration)}
\affiliation{Department of Physics, Carnegie Mellon University, 
             Pittsburgh, PA 15213, USA}

\begin{abstract}
Progress in extracting excited-state baryon masses
in lattice QCD using large sets of spatially-extended operators
is presented.  The use of stochastic estimates of all-to-all
quark propagators with variance reduction techniques is described.
Such techniques are crucial for incorporating multi-hadron
operators into the correlation matrices of the hadron operators.
\end{abstract}

\maketitle

\thispagestyle{fancy}


\section{INTRODUCTION}
Experiments show that many excited-state hadrons exist, and there are significant
experimental efforts to map out the QCD resonance spectrum, such as Hall B
and the proposed Hall D at Jefferson Lab, ELSA associated with the University
of Bonn, COMPASS at CERN, PANDA at GSI, and BESIII in Beijing.
Hence, there is a great need for \textit{ab initio} determinations of such
states in lattice QCD.

To extract excited-state energies in Monte Carlo calculations, 
correlation matrices are needed and operators with very good overlaps onto 
the states of interest are crucial.  
To study a particular state of interest, all states lying below that
state must first be extracted, and as the pion gets lighter in lattice QCD 
simulations, more and more multi-hadron states will lie below the excited
resonances.  To reliably extract these multi-hadron states, multi-hadron 
operators made from constituent hadron operators with well-defined relative
momenta will most likely be needed, and the computation of temporal correlation
functions involving such operators will require the use of all-to-all quark 
propagators.  The evaluation of disconnected diagrams will ultimately be
required.  Perhaps most worrisome, most excited hadrons are unstable 
(resonances), so the results obtained for finite-box stationary-state
energies in lattice QCD must be interpreted carefully.

In this talk, progress by the Hadron Spectrum Collaboration 
in extracting excited-state baryon masses
in lattice QCD using large sets of spatially-extended operators
is presented.  The use of stochastic estimates of all-to-all
quark propagators with variance reduction techniques is described.
Such techniques are crucial for incorporating multi-hadron
operators into correlation matrices. 
\vspace*{-4mm}

\section{EXCITED STATIONARY STATES IN LATTICE QCD: KEY ISSUES}

Capturing the masses of excited states requires the computation
of correlation matrices $C_{ij}(t)=\langle 0\vert T \Phi_i(t)
 \Phi^\dagger_j(0) \vert 0\rangle$ associated with a large set of $N$
different operators $\Phi_i(t)$.  It has been shown in Ref.~\cite{wolff90}
that the $N$ {\em principal effective masses} $W_\alpha(t)$, defined by
\[  W_\alpha(t)=\ln\left(\frac{\lambda_\alpha(t,t_0)}{
 \lambda_\alpha(t+1,t_0)}\right),
\]
where $\lambda_\alpha(t,t_0)$ are the eigenvalues of
 $C(t_0)^{-1/2}\ C(t)\ C(t_0)^{-1/2}$ and $t_0<t/2$ is usually chosen,
tend to the eigenenergies of the lowest $N$ states with which the
$N$ operators overlap as $t$ becomes large.   When combined
with appropriate analysis methods, such variational
techniques are a particularly powerful tool for investigating excitation
spectra. 

\begin{figure}[t]
\centerline{
\raisebox{0mm}{\setlength{\unitlength}{1mm}
\thicklines
\begin{picture}(16,10)
\put(8,6.5){\circle{6}}
\put(7,6){\circle*{2}}
\put(9,6){\circle*{2}}
\put(8,8){\circle*{2}}
\put(4,0){single-}
\put(5,-3){site}
\end{picture}}
\hspace*{-7mm}
\raisebox{0mm}{\setlength{\unitlength}{1mm}
\thicklines
\begin{picture}(16,10)
\put(7,6.2){\circle{5}}
\put(7,5){\circle*{2}}
\put(7,7.3){\circle*{2}}
\put(14,6){\circle*{2}}
\put(9.5,6){\line(1,0){4}}
\put(4,0){singly-}
\put(2,-3){displaced}
\end{picture}}
\hspace*{-7mm}
\raisebox{0mm}{\setlength{\unitlength}{1mm}
\thicklines
\begin{picture}(20,8)
\put(12,5){\circle{3}}
\put(12,5){\circle*{2}}
\put(6,5){\circle*{2}}
\put(18,5){\circle*{2}}
\put(6,5){\line(1,0){4.2}}
\put(18,5){\line(-1,0){4.2}}
\put(6,0){doubly-}
\put(4,-3){displaced-I}
\end{picture}}
\hspace*{-4mm}
\raisebox{0mm}{\setlength{\unitlength}{1mm}
\thicklines
\begin{picture}(20,13)
\put(8,5){\circle{3}}
\put(8,5){\circle*{2}}
\put(8,11){\circle*{2}}
\put(14,5){\circle*{2}}
\put(14,5){\line(-1,0){4.2}}
\put(8,11){\line(0,-1){4.2}}
\put(4,0){doubly-}
\put(1,-3){displaced-L}
\end{picture}}
\hspace*{-6mm}
\raisebox{0mm}{\setlength{\unitlength}{1mm}
\thicklines
\begin{picture}(20,12)
\put(10,10){\circle{2}}
\put(4,10){\circle*{2}}
\put(16,10){\circle*{2}}
\put(10,4){\circle*{2}}
\put(4,10){\line(1,0){5}}
\put(16,10){\line(-1,0){5}}
\put(10,4){\line(0,1){5}}
\put(4,0){triply-}
\put(1,-3){displaced-T}
\end{picture}}
\hspace*{-6mm}
\raisebox{0mm}{\setlength{\unitlength}{1mm}
\thicklines
\begin{picture}(20,12)
\put(10,10){\circle{2}}
\put(6,6){\circle*{2}}
\put(16,10){\circle*{2}}
\put(10,4){\circle*{2}}
\put(6,6){\line(1,1){3.6}}
\put(16,10){\line(-1,0){5}}
\put(10,4){\line(0,1){5}}
\put(4,0){triply-}
\put(2,-3){displaced-O}
\end{picture}} 
\hspace*{-7mm}
\raisebox{0mm}{\setlength{\unitlength}{1mm}
\thicklines
\begin{picture}(20,12)
\put(7,7){\circle{2}}
\put(9,7){\circle*{2.5}}
\put(4,0){single-}
\put(5,-3){site}
\end{picture}}
\hspace*{-11mm}
\raisebox{0mm}{\setlength{\unitlength}{1mm}
\thicklines
\begin{picture}(20,12)
\put(6,7){\circle{2}}
\put(12,7){\circle*{2.5}}
\put(7,7){\line(1,0){4}}
\put(4,0){singly-}
\put(2,-3){displaced}
\end{picture}} 
\hspace*{-8mm}
\raisebox{0mm}{\setlength{\unitlength}{1mm}
\thicklines
\begin{picture}(20,12)
\put(7,11){\circle{2}}
\put(13,5){\circle*{2.5}}
\put(12,5){\line(-1,0){5}}
\put(7,10){\line(0,-1){5}}
\put(4,0){doubly-}
\put(1,-3){displaced-L}
\end{picture}}
\hspace*{-6mm}
\raisebox{0mm}{\setlength{\unitlength}{1mm}
\thicklines
\begin{picture}(20,12)
\put(8,11){\circle{2}}
\put(14,11){\circle*{2.5}}
\put(8,4){\line(1,0){6}}
\put(14,4){\line(0,1){6}}
\put(8,4){\line(0,1){6}}
\put(4,0){triply-}
\put(2,-3){displaced-U}
\end{picture}}
\hspace*{-6mm}
\raisebox{0mm}{\setlength{\unitlength}{1mm}
\thicklines
\begin{picture}(20,15)
\put(7,5){\circle{2}}
\put(11,7){\circle*{2.5}}
\put(7,12){\line(1,0){8}}
\put(7,6){\line(0,1){6}}
\put(15,12){\line(-3,-4){3.0}}
\put(4,0){triply-}
\put(2,-3){displaced-O}
\end{picture}}
}
\vspace*{8pt}
\caption{The spatial arrangements of the extended three-quark baryon
and quark-antiquark meson operators. Solid circles are smeared quark fields,
open circles in the mesons are smeared ``barred'' antiquark fields,
open circles in the baryons indicate a Levi-Civita color coupling,
and solid line segments indicate covariant displacements.
\label{fig:operators}}
\end{figure}
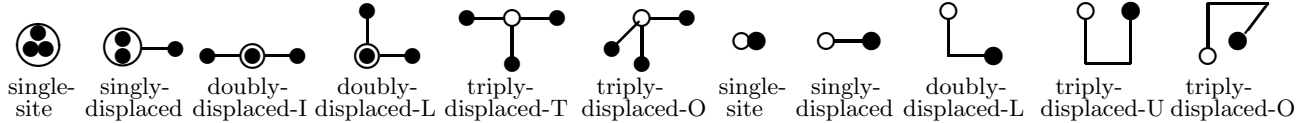

The use of operators whose correlation functions $C(t)$ attain their
asymptotic form as quickly as possible is crucial for reliably
extracting excited hadron masses.  An important ingredient in constructing
such hadron operators is the use of smeared fields.  Operators constructed
from smeared fields have dramatically reduced mixings with the high frequency
modes of the theory.  Both link-smearing and quark-field smearing should
be applied.  Since excited hadrons are expected to be large objects, 
the use of spatially extended operators is another key ingredient in
the operator design and implementation.  Fig.~\ref{fig:operators} shows
the different spatial configurations we use, which effectively
build up the necessary orbital and radial structures of the hadron
excitations.  The basic building blocks in all of our hadron operators
are covariantly-displaced quark or antiquark fields. These are first
combined to have the appropriate flavor structure and color structure,
then group-theoretical projections are applied to obtain operators which
transform irreducibly under all lattice rotation and reflection symmetries.
A more detailed discussion of these issues can be found in Ref.~\cite{baryons1}.

\begin{figure}
\begin{center}
\includegraphics[width=2.0in,bb=6 94 532 662]{nucleon_spectrum_lat.eps}
\hspace*{10mm}
\includegraphics[width=2.0in,bb=156 260 421 555]{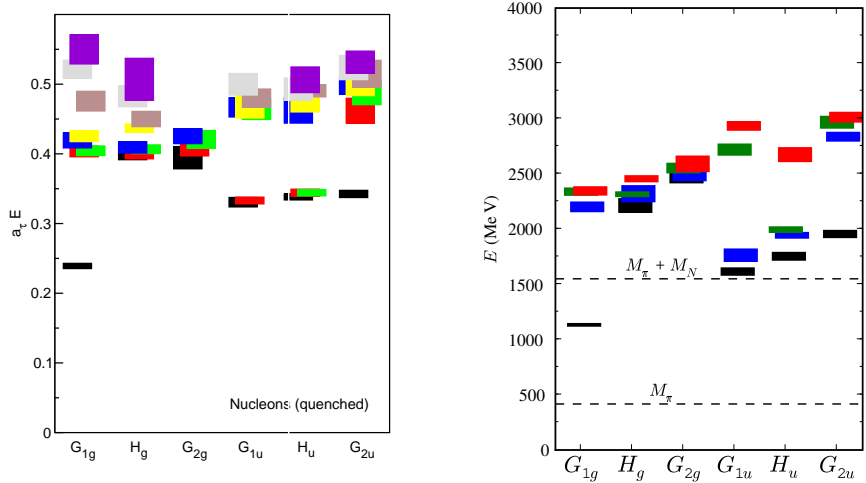}
\end{center}
\caption[nucleons]{
(Left) Nucleon spectrum from 200 quenched configurations
on a $12^3\times 48$ anisotropic lattice using the Wilson gauge and quark
actions with $a_s \sim 0.1$ fm,
$a_s/a_t \sim 3.0$ and $m_\pi\sim 700$~MeV from Ref.~\cite{adamthesis}.  
(Right) Nucleon spectrum from 430 $N_f=2$ configurations on a $24^3\times 64$
lattice using a stout-smeared clover fermion action and Symanzik-improved
gauge action with $a_s\sim 0.1$~fm, $a_s/a_t\sim 3$, and $m_\pi=400$~MeV
from Ref.~\cite{nucleons2}.
\label{fig:nucleonspectra}}
\end{figure}

A first glimpse of the higher-lying nucleon spectrum in lattice QCD was
provided by the Hadron Spectrum Collaboration in Ref.~\cite{adamthesis}.
These first results, shown in Fig.~\ref{fig:nucleonspectra}, were on small
$12^3\times 48$ anisotropic quenched lattices with a very heavy pion.
Results for both the nucleons and $\Delta$-resonances on
239 quenched configurations on a $16^3\times 64$ lattice and 167 quenched
configurations on a $24^3\times 64$ lattice using an anisotropic Wilson
action with spatial spacing $a_s\sim 0.1$~fm, $a_s/a_t\sim 3$, and a
pion mass $m_\pi\sim 490$~MeV appeared during the past 
year\cite{nucleons1}.  These masses have also been determined in the
past year using 430 $N_f=2$ configurations on a $24^3\times 64$ 
lattice with a stout-smeared 
clover fermion action and a Symanzik-improved anisotropic gauge 
action\cite{nucleons2}.  The results for a pion mass $m_\pi=400$~MeV,
spacing $a_s\sim 0.1$~fm and $a_s/a_t\sim 3$
are shown in Fig.~\ref{fig:nucleonspectra}.  The low-lying odd-parity
band shows the exact number of states in each channel as expected
from experiment.  The two figures show the splittings in the band
increasing as the quark mass is decreased.  At these heavy pion masses,
the first excited state in the $G_{1g}$ channel is significantly higher
than the experimentally measured Roper resonance.  It remains to be
seen whether or not this level will drop down as the pion mass is further
decreased.  Most of the levels in the right-hand plot lie very close
to two-particle thresholds.  The use of two-hadron operators will be
needed to go to lighter pion masses.

\section{MANY-TO-MANY QUARK PROPAGATORS}

To study a particular eigenstate of interest, all eigenstates lying below that
state must first be extracted, and as the pion gets lighter in lattice QCD 
simulations, more and more multi-hadron states will lie below the excited
resonances.  The correlation functions of such operators require
estimates of the quark propagators from all spatial sites on a time slice 
to all spatial sites on another time slice.   Computing all such elements
of the propagators exactly is not possible (except on very small lattices).
Some way of stochastically estimating them is needed.

Random noise vectors $\bm{\eta}$ whose expectations satisfy
$E(\eta_i)=0$ and $E(\eta_i\eta_j^\ast)=\delta_{ij}$ are useful for 
stochastically estimating the inverse of a large matrix $M$ 
as follows.  Assume that for each of $N_R$ 
noise vectors, we can solve the following
linear system of equations: $M X^{(r)}=\eta^{(r)}$ for $X^{(r)}$.
Then $X^{(r)}=M^{-1}\eta^{(r)}$, and
\begin{equation}
   E( X_i \eta_j^\ast ) = E( \sum_k M^{-1}_{ik}\eta_k \eta_j^\ast )
  = \sum_k M^{-1}_{ik}
  E(\eta_k \eta_j^\ast) = \sum_k M^{-1}_{ik} \delta_{kj} = M^{-1}_{ij}.
\end{equation}
The expectation value on the left-hand can be approximated using the
Monte Carlo method.  Hence, a Monte Carlo estimate of $M_{ij}^{-1}$
is given by
\begin{equation}
  M_{ij}^{-1} \approx \lim_{N_R\rightarrow\infty}\frac{1}{N_R}
 \sum_{r=1}^{N_R} X_i^{(r)}\eta_j^{(r)\ast}, \qquad
\mbox{where $MX^{(r)}=\eta^{(r)}$.}
\end{equation}
This equation usually produces stochastic 
estimates with variances which are much too large to be useful.

Progress is only possible if stochastic estimates of the quark
propagators with reduced variances can be made.  A technique of
\textit{diluting} the noise vectors has been developed which
accomplishes such a variance reduction\cite{dilute1}.
A given dilution scheme can be viewed as the application of a complete
set of projection operators.  To see how dilution works, consider a general 
$N\times N$ matrix $M$ having matrix elements $M_{ij}$.  
Define some complete set of $N\times N$ projection matrices $P^{(a)}$ which 
satisfy
$
 P^{(a)}P^{(b)}=\delta^{ab}P^{(a)}$ with $\sum_a P^{(a)}=1$
and $P^{(a)\dagger}=P^{(a)}$.  Then observe that
\begin{eqnarray}
  M_{ij}^{-1}&=&M_{ik}^{-1}\delta_{kj}=\sum_a M_{ik}^{-1}P^{(a)}_{kj}
  =\sum_a M_{ik}^{-1}P^{(a)}_{kk^\prime}P^{(a)}_{k^\prime j}
  =\sum_a M_{ik}^{-1}P^{(a)}_{kk^\prime}\delta_{k^\prime j^\prime} P^{(a)}_{j^\prime j}
 \nonumber \\
  &=&\sum_a M_{ik}^{-1}P^{(a)}_{kk^\prime}E(\eta_{k^\prime}\eta^\ast_{j^\prime}) 
    P^{(a)}_{j^\prime j}
   =\sum_a M_{ik}^{-1}E\left(P^{(a)}_{kk^\prime}\eta_{k^\prime}
    \eta^\ast_{j^\prime}P^{(a)}_{j^\prime j}\right). 
\end{eqnarray}
Define
$
  \eta^{[a]}_k=P^{(a)}_{kk^\prime}\eta_{k^\prime}$ and
 $ \eta^{[a]\ast}_j=\eta^\ast_{j^\prime}P^{(a)}_{j^\prime j}
    =P^{(a)\ast}_{j j^\prime}\eta^\ast_{j^\prime},
$
and further define $X^{[a]}$ as the solution of
$
   M_{ik}X^{[a]}_k=\eta^{[a]}_i,
$
then we have
\begin{equation}
   M_{ij}^{-1}=\sum_a M_{ik}^{-1} E(\eta^{[a]}_k \eta^{[a]\ast}_j)
      =\sum_a E(X^{[a]}_i\eta^{[a]\ast}_j).
\label{eq:diluted}
\end{equation}
Although the expected value of $\sum_a \eta^{[a]}_k
\eta^{[a]\ast}_j$ is the same as $\eta_k\eta^\ast_j$, the \textit{variance} 
of $\sum_a \eta^{[a]}_k\eta^{[a]\ast}_j$ is significantly smaller than that 
of $\eta_k\eta^\ast_j$.  For both $Z_4$ and $U(1)$ noise, we
have
$  Var({\rm Re}(\eta_i\eta_j^\ast))=Var({\rm Im}(\eta_i\eta_j^\ast))=
 \textstyle\frac{1}{2}(1-\delta_{ij}).$
Although the variance is zero for $i=j$, there is a significant variance 
for all $i\neq j$.  The dilution projections ensure \textit{exact zeros} 
for many of the off-diagonal elements, instead of values that are only
statistically zero.  In other words, many of the $i\neq j$ elements
become exactly zero.

Of course, the effectiveness of the variance reduction depends on the
projectors chosen.  A particularly important dilution scheme for measuring 
temporal correlations in hadronic quantities is ``time dilution'' where 
the noise vector is broken up into pieces which only have support on a
single time slice.  Spin and color dilution are two other easy-to-implement
schemes, and various spatial dilution schemes are possible.  These various 
dilution projectors can also be combined to make hybrid schemes.

\begin{figure}
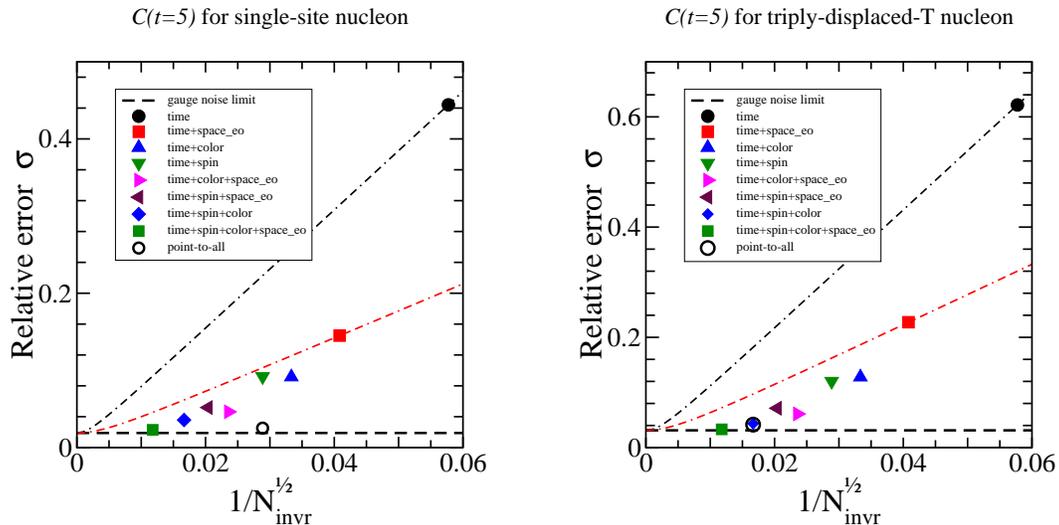

\begin{center}
\includegraphics[width=2.5in,bb=17 55 547 581]{SS_C5_dilutions.eps}
\hspace*{10mm}
\includegraphics[width=2.5in,bb=17 55 547 581]{TDT_C5_dilutions.eps}
\end{center}
\caption[dilute]{
(Left) The relative errors in the correlation function of a single-site 
nucleon operator for temporal separation $t=5a_t$ evaluated using
stochastically-estimated quark propagators with different dilution
schemes against $1/N_{\rm inv}^{1/2}$, where $N_{\rm inv}$ is the
number of Dirac matrix inversions required. The open circle shows the
point-to-all error, and the horizontal dashed line shows the gauge-noise
limit.  The black (red) dashed-dotted line shows the decrease in error
expected by simply increasing the number of noise vectors, starting 
from the time (time + even/odd-space) dilution point.  (Right)
Same as the left plot, except for a triply-displaced-T nucleon operator.
These results used 100 quenched configurations on a $12^3\times 48$
lattice.
\label{fig:dilutions}}
\end{figure}

A comparison of the different dilution schemes is shown in
Fig.~\ref{fig:dilutions}.  These computations are dominated by the 
inversions of the Dirac matrix, so using the number of matrix inversions
$N_{\rm inv}$ to compare computational efforts is reasonably fair. 
The advantage in using increased dilutions
compared to an increased number of noise vectors with only time dilution
is evident in the plots.  However, this advantage quickly diminishes after
time + even/odd-space dilution, or time+color, or time+spin dilution.
These encouraging results demonstrate that the inclusion of good multi-hadron
operators will certainly be possible using stochastic all-to-all
quark propagators with diluted-source variance reduction. 

\section{SUMMARY AND OUTLOOK}

This talk discussed the key issues and challenges in exploring excited
hadrons in lattice QCD. The importance of multi-hadron operators
and the need for all-to-all quark propagators were emphasized.  The technology
needed to extract excited stationary-state energies, including operator 
design and field smearing, was detailed.  Efforts in variance reduction of
stochastically-estimated all-to-all quark propagators using source dilutions
was described.

Given the major experimental efforts to map out the QCD resonance spectrum, 
such as Hall B and the proposed Hall D at Jefferson Lab, ELSA, COMPASS, 
PANDA, and BESIII,
there is a great need for \textit{ab initio} determinations of such
states in lattice QCD.
The exploration of excited hadrons in lattice QCD is well underway.
\vspace*{-7mm}

\begin{acknowledgments}
Members of the Hadron Spectrum Collaboration are 
John Bulava, Saul Cohen, Jozef Dudek, Robert Edwards, Eric Engelson, 
Justin Foley, Balint Joo, Jimmy Juge, Huey-Wen Lin, Nilmani Mathur,
Mike Peardon, David Richards, Sinead Ryan, and Steve Wallace.
This work was supported by the National Science Foundation through awards
PHY 0653315 and PHY 0510020.
\end{acknowledgments}

\end{document}